\documentclass[11pt]{article}
\usepackage{fleqn,cospar}

\usepackage{url}

\usepackage{epsfig}
\usepackage[figuresright]{rotating}

\def\zycki{$\dot{\rm Z}$ycki}
\def\rozanska{R\^o$\dot{\rm z}$a\'nska}
\def\lubinski{Lubi\'nski}
\def\gierlinski{Gierli\'nski}

\title{GALACTIC BLACK HOLE BINARY SYSTEMS}

\author{C. Done\address{Department of Physics, 
    University of Durham, South Road, Durham, DH1 3LE, UK}}

\begin{document}

\maketitle

\begin{abstract}

I review observations of the X--ray spectra of Galactic Black Hole
Candidates, and theoretical ideas as to how these can be produced.
X--ray reflection should enable different source geometries to be
distinguished, but the ionisation instability of X--ray irradiated
material in hydrostatic equilibrium gives rise to large ambiguities in
interpretation. It is not currently possible to determine whether the
underlying emission mechanism in the low state is an advective flow or
magnetic reconnection above the disk, but more detailed modelling of
the ionisation instability {\it may} allow us to distinguish between
these scenarios. 

\end{abstract}

\section*{INTRODUCTION}

Accretion onto a black hole is observed to produce copious X--ray
emission.  At high mass accretion rates (approaching the Eddington
limit) the
spectra are dominated by a soft component at $kT\sim 1$ keV which is
strongly (very high state: VHS) or weakly (high state: HS) Comptonized
by low temperature thermal (or quasi--thermal) electrons with $kT\sim
5-20 $ keV (e.g. \gierlinski\ et al., 1999). There is also a rather
steep power law tail ($\Gamma\sim 2-3$) which extends out beyond $511$
keV in the few objects with good high energy data (e.g. Grove et al.,
1998). At lower mass accretion rates, below $\sim 2-3$ per cent of
Eddington (\zycki\ et al., 1998; \gierlinski\ et al., 1999), there is
a rather abrupt transition when the soft component drops in
temperature and luminosity. Instead this (low state: LS) spectrum is
dominated by thermal Comptonization, with $\Gamma < 1.9$, rolling over
at energies of $\sim 150$ keV (see e.g. the reviews by Tanaka \& Lewin
1995; van der Klis 1995; Nowak 1995).  This spectral form seems to
continue even down to very low luminosities ($\sim 0.01$ per cent of
Eddington: the quiescent or off state e.g. Kong et al. 2000).  These
very different spectra can all be seen from a single object where the
mass accretion rate changes substantially.  Figure 1 (left panel) 
shows VHS, HS and
LS spectra from the GBHC transient RXTE J1550-564.

\begin{figure}
\begin{tabular}{cc}
{\epsfig{file=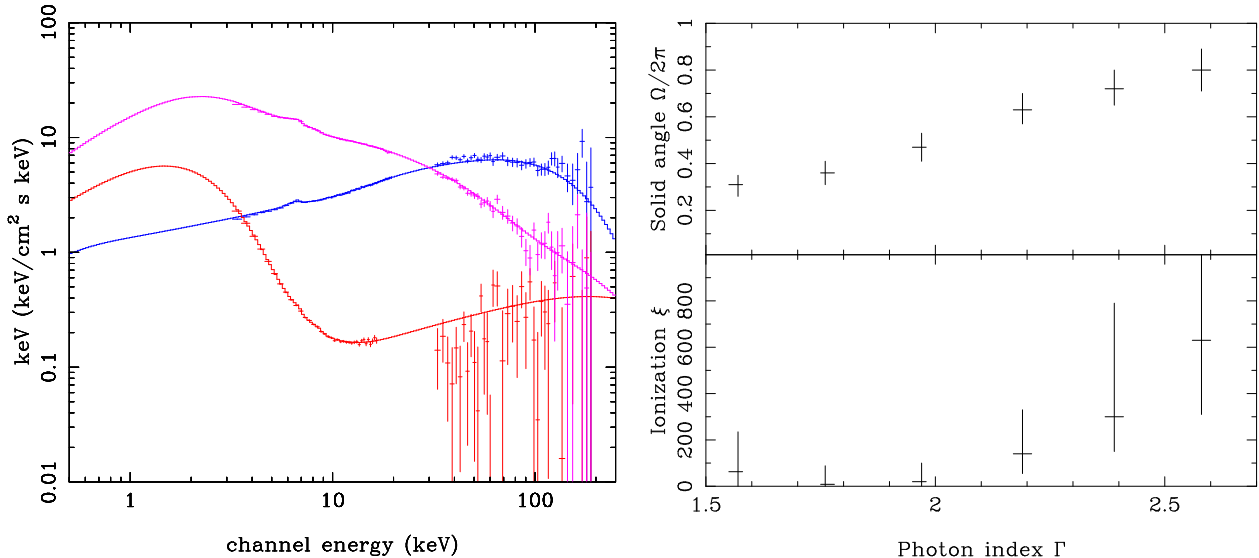,width=0.45\hsize}} &
{\epsfig{file=done_fig1b.ps,width=0.45\hsize}}
\end{tabular}
\vskip -24pt
\caption{Left panel: RXTE PCA and HEXTE data from RXTE J1550--564 showing the
Low state (peaking at 100 keV), High
state (sharply peaking at 1 keV) and very high state (smooth peak at 1
keV). Right panel: Fixed density reflection models fit to simulated spectra
calculated from X--ray illuminated disk models including the 
self--consistent density structure (NKK). The ionisation instability
forms an ionised skin on top of the disk which is completely
reflective for flat spectrum illumination (high Compton temperature),
so suppressing the observed reflected signature.
For steeper spectra the skin depth decreases (from Done \& Nayakshin
2000). }  
\end{figure}

Intriguingly, other accreting systems show the same sort of
hard--to--soft spectral switch. The low magnetic field neutron star
systems (atolls) show such a tranistion at $\sim 0.8-1$ per cent of
Eddington (Ford et al., 2000), i.e. a factor of 2--3 lower than the
GBHC. In AGN, nearby Seyfert galaxies generally have hard X--ray
spectra which are rather similar to the LS GBHC spectra (Zdziarski et
al 1995), while the (probably higher mass accretion rate) Narrow Line
Seyfert 1 galaxies and radio quiet quasars look more similar to the
HS/VHS spectra (e.g. the review by \zycki\ 2000, these proceedings).

\section*{HARD X--RAY EMISSION MECHANISMS}

The standard accretion disk models (Shakura \& Sunyaev 1973; SS)
cannot produce hard X--ray emission. Some additional process is
required where a large fraction of the gravitational energy released
by accretion is dissipated in an optically thin environment.  For the
SS disk, an obvious candidate is that there are magnetic flares above
the disk, generated by the Balbus--Hawley MHD dynamo responsible for
the disk viscosity (e.g. the review by Balbus \& Hawley
1998). Buoyancy could cause the magnetic field loops to rise up to the
surface of the disk, so they can reconnect in regions of fairly low
particle density. Numerical simulations (albeit incomplete) do show
this happening, although they do not yet carry enough
power to reproduce the observed low/hard state (Miller \& Stone 2000).

Such mechanisms change the SS disk structure, as much of the power is
dissipated in a magnetic corona rather than in a disk (which is then
cooler and denser: Svensson \& Zdziarski 1994). However, the strong
illumination of the disk by the corona also changes its structure to
some (as yet not well understood) extent.  Little is known about the
spectrum expected from electrons heated through magnetic reconnection,
but the observed weak disk emission in the LS can be explained if most
of the viscous dissipation is released in a patchy or outflowing
magnetic corona (Svensson \& Zdziarski 1994; Stern et al 1995;
Beloborodov 1999). Thus it seems likely that such mechanisms could
work to form the LS hard X--ray emission. However, there could be
considerable problems in extending these ideas to the much lower
luminosity quiescent X--ray state as numerical simulations of the MHD
turbulence imply that the magnetic field shuts off when the disk
material becomes mainly neutral as in quiescence (Gammie \&
Menou 1998; Fleming et al., 2000).

An alternative to the SS disk models is that the inner disk is
replaced by an optically thin, X--ray hot accretion flow. By contrast
with the magnetic reconnection models, these have spectra which can be
worked out in some detail, typically giving electron temperatures of
$\sim 100$ keV (advection dominated accretion flows: Narayan \& Yi
1995). These can exist only at fairly low mass accretion rates as they
rely crucially on the assumptions that the accretion energy is given
mainly to the protons, and that the electrons are only heated via
Coulomb collisions. At high densities (i.e. high mass accretion rates)
the electrons efficiently drain energy from the protons and the
flow collapses back into an SS disk at luminosities of $\sim 
\alpha_{0.1}^2$ per cent of Eddington (Esin et al., 
1997; Quataert \& Narayan
1999). These flows are also stable to convection (Narayan, Igumenshchev \&
Abramowicz 2000) at the high viscosity ($\alpha\sim 0.2$) 
required to get the LS--HS/VHS tranistion at the observed luminosities.
Hence these flows could produce both the quiescent
and LS spectra, and a physical
mechanism for the HS/VHS state transition (Esin et al.,  1997). 

Magnetic reconnection is really the only known mechanism that can give
the power law tail seen in the HS/VHS. There
is also a radio jet/outflow seen in the VHS (and LS) but this is
strongly quenched in HS (e.g. Fender 2000) while the hard X--ray tail
can stay relatively constant (e.g. left panel of Figure 1 ).  An alternative
possibility is that the hard X--ray tail comes from the accelerating
infall between the last stable orbit and the event horizon. Compton
scattering from the bulk motion of the accelerating flow can upscatter
any seed photons intercepted by this material into a power law
tail. However, the free--fall electron velocities are not high enough
(typical velocities of $\sim 0.7c$ imply Lorenz factors of only
$\gamma\sim 1.4$) to extend the power law past $\sim 100-300$ keV
(e.g. Laurent \& Titarchuk 1999, Zdziarski 2000) yet the highest
signal--to--noise HS/VHS spectra extend unbroken beyond this
(e.g. Grove et al 1998). 

There is also some thermal Comptonization of the disk photons in the
HS and (most strongly) VHS (\gierlinski\ et al., 1999, Wilson \& Done
2000). This could be due to heated upper layers of 
the inner disk, Comptonizing the radiation from beneath as it
escapes (e.g. \zycki\ et al., 1999b). Alternatively, 
magnetic reconnection above the disk could produce both the
non--thermal and themal electron distributions (hybrid plasma:
Poutanen \& Coppi 1998; \gierlinski\ et al., 1999).

Figure 2 sketches all these potential hard X--ray emission mechanisms
for each state. 

\begin{figure}
\begin{tabular}{c}
\epsfig{file=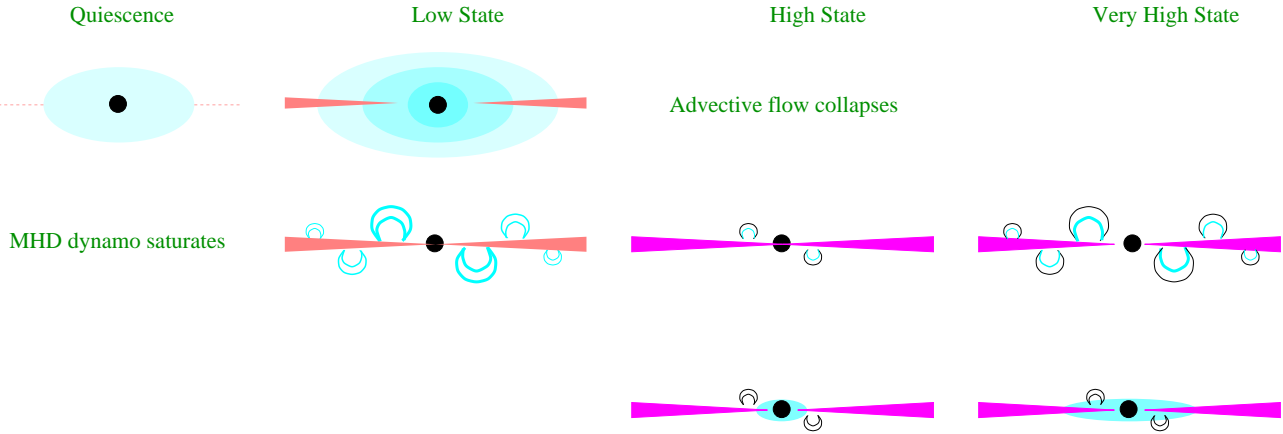,width=0.95\hsize}
\end{tabular}
\vskip -24pt
\caption{Potential X--ray emission mechanisms in the various spectral
states. In quiescence the disk is NOT in steady state (as indicated by
the dotted line). Hydrogen is mostly neutral and the MHD dynamo
probably cannot operate. In the low state the accretion flow is in 
(quasi)steady state. Hydrogen is ionised so the MHD dynamo works, and
if there is reconnection above the disk then this could heat the electrons.
Alternatively, the X--ray emission in 
both quiescence and and low state could be powered by an advective
flow. In the high and very high states the only serious contender for
the hard power law tail is magnetic reconnection leading to a 
non--thermal electron distribution (indicated by the black loops),
while the thermal electrons could be part of the same mechanism (grey
loops) or could be associated with the inner disk.}
\end{figure}

\section*{TESTING THE MODELS}

These very different mechanisms for the hard X--ray emission also
imply very different geometries (see Figure 2). If there was a way to
constrain the hard X--ray source geometry then we could start to
distinguish between the emission mechanisms. The overall continuum
shape gives the first constraints as the Comptonised spectrum is
determined by the ratio of power in the seed photons to that in the
hard electrons (e.g. Haardt \& Maraschi 1993; Poutanen et al 1997).

\subsection*{Variability Behaviour}

Another potential diagnostic is from the variability behaviour of the
source, since their power spectra show characteristic frequencies in
the form both of breaks and Quasi Periodic Oscillations (QPO's). These
features are related ($f_{break}\sim 10 f_{QPO}$ for the low frequency
QPO), and they {\it move}, with the frequencies generally being higher
(indicating smaller size scales) as the source changes from LS to VHS
(see e.g. the review by van der Klis 2000). However, the physical
origin of these features are not well understood, and there are
several models for QPO formation (see e.g. the review by van der Klis
2000).  Recent progress has concentrated on the similarity between the
relationship between the QPO and break frequencies in black holes {\it
and} neutron star systems (Psaltis, Belloni \& van der Klis 1999). If
they truly are the same phenomena then the mechanism {\it must} be
connected to the accretion disk properties and not to the
magnetosphere or surface of the neutron star. One very promising
approach is identifying the (low frequency) QPO's with the vertical
precession timescale of a tilted ring of gas at the inner radius of
the disk (Stella \& Vietri 1998). This model was extended by Psaltis
\& Norman (2000), who calculated the transfer function of the disk to
a spectrum of perturbations, showing that the disk acts as a low pass
filter, with the hydrodynamic timescale at the inner edge suppressing
fast variability (and so producing a low frequency break), with a
superimposed QPO at the Lense--Thirring precession timescale.  While
this is tantalisingly close to observations, there are still some
significant problems -- most importantly connected to the neutron star
spin frequencies which may have been measured during X--ray bursts
(e.g. van der Klis 2000). 
However, despite these uncertainties, {\it all} QPO and break
frequency models use a sharp transition in the accretion disk in some
form to pick out a preferred timescale, so this should give {\it some}
constraints on the geometry.

\subsection*{Reflected Spectra}

Yet another, independent test of the geometry comes from the X--ray
spectroscopy. Wherever hard X--rays illuminate optically thick
material then a reflected spectrum is produced which has a
characteristic spectral shape, with an especially prominant iron
K$\alpha$ iron fluorescence line. The amount of line and reflected
continuum depend on the solid angle subtended by the material, the
ionization state and elemental abundances of the reflector (Lightman
\& White 1988; George \& Fabian 1991; Matt, Perola \& Piro 1991; Ross
\& Fabian 1993; \zycki\ \& Czerny 1994).

The reflected spectrum (line and continuum) is smeared by special and
general relativistic effects of the motion of the disk in the deep
gravitational potential well (Fabian et al., 1989). The fraction of
the incident flux which is reflected gives a measure of the solid
angle of the optically thick disk as observed from the hard X--ray
source, while the amount of smearing shows how far the material
extends into the gravitational potential of the black hole.

\begin{figure}
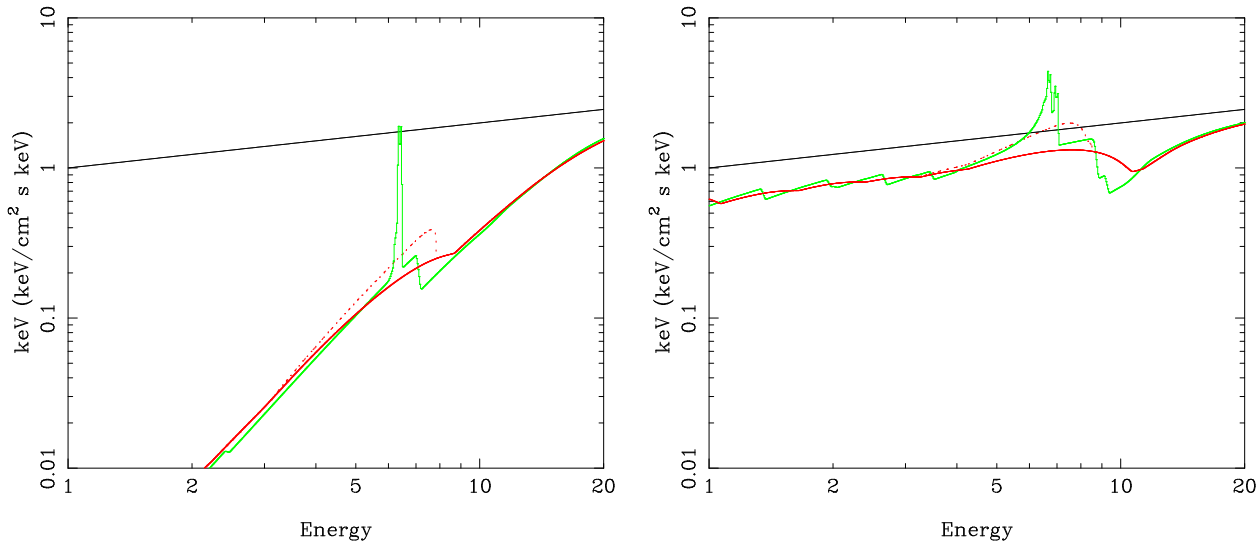

\begin{tabular}{cc}
{\epsfig{file=done_fig3a.ps,width=0.45\hsize}} &
{\epsfig{file=done_fig3b.ps,width=0.45\hsize}} \\
\end{tabular}
\vskip -24pt
\caption{Left panel: The light grey (green) 
line shows the $\nu f(\nu)$ spectrum reflected from a slab
of neutral material of solar abundance viewed at $60^\circ$. The dark
grey (red) line shows the same spectrum but from a disk extending down to
the last stable orbit around a black hole (solid line is continuum
only, while dotted line is continuum plus line). The black line is the
illuminating spectrum. Right panel: As in (a) but for highly ionised material}
\end{figure}

The left panel of Figure 3 shows the reflected spectrum from a slab of
neutral, solar abundance material viewed at $60^{\circ}$ (light grey/green)
compared to that from a disk extending down to 3 Schwarzchild radii
around a black hole. The Greens functions for the reflected continuum
are taken from Magdziarz \& Zdziarski (1995), which include Compton
downscattering at all energies. The iron line is calculated self
consistently with the continuum reflection (including Compton
downscattering on the line profile) using the code of \zycki\ \&
Czerny (1994), and the whole spectrum (line and continuum) is smeared
assuming that the flux irradiating the accretion disk $F_{irr}\propto
r^{-3}$ (see Appendix A of \zycki\ et al., 1999).

The right panel of Figure 3
shows the same for a highly ionised reflector. The resulting
spectrum now depends strongly on details of the assumed ionization
structure, both as a function of height and of radius. Here we assume that
the slab has a constant ionization parameter $\xi=L/(nr^2)$ (where $L$
is the illuminating luminosity, at a distance $r$ from the reflector
of constant density $n$). The photo--electric opacity is calculated by
determining the ion populations (balancing photo--ionization with
collisional recombination as as in Done et al., 1992). The electron
temperature may then cause appreciable Compton upscattering (Ross \&
Fabian 1993; Ross, Fabian \& Young 1999). This is
{\it not} included in the reflected continuum
calculation (Magdziarz \& Zdziarski 1995), although it is included to
some extent in the line profile (\zycki\ \& Czerny 1994 but see the
discussion in Ross et al., 1999).

As can be seen from these figures, the fraction of reflected flux in
the 1--20 keV band is dependent on the ionization state of the
material as well as the solid angle subtended by it. The ionization
state is determined mostly by the line and edge energies (although the
reflection continuum shape is also important). But these energies can
be shifted by relativistic effects (Ross, Fabian \& Brandt 1996). 
Thus all three parameters
(solid angle, ionization state and relativistic smearing) should be
fit simultaneously in order to determine the
geometry of the system.

\section*{COMPARISON WITH OBSERVATIONS OF GBHC: QUIESCENCE AND LOW STATE }

The quiescent spectra are too faint to do detailed spectral fitting or
variability power spectra, but they do appear to be hard $\Gamma < 2$
i.e. to be related to the LS spectra (e.g. 
Kong et al., 2000). There is much more information on the (much
brighter) LS spectra. 

\subsection*{Continuous static magnetic corona}

The LS spectra are relatively hard $\sim 1.5 < \Gamma < 1.9$ extending
out to $\sim 100-200$ keV, which is an immediate problem for a
`disk--corona' geometry where the magnetic flares are close enough
together to form a quasi--continuous corona above the disk.  In such a
geometry, the hard X--ray corona radiates about as much power down
towards the disk as up towards the observer. The downwards flux
illuminates the disk, and some fraction (given by the energy
integrated reflection albedo) is reflected but the rest is thermalised
in the disk, forming an additional seed photon source (Haardt \&
Maraschi 1993).  The reflection albedo cannot approach unity even if
the disk is completely ionised due to Compton downscattering of the
high energy spectrum. For $\Gamma=1.5$ the
maximum albedo $\sim 0.5$, so the {\it minimum} soft seed flux is
approximately a quarter of the hard luminosity. Yet this ratio of
soft--to--hard flux would result in a Compton spectrum with
$\Gamma\sim 1.8-1.9$ (Haardt \& Maraschi 1993; Poutanen 1998).  
Thus the `disk--corona' geometry is ruled
out. If the magnetic corona exists, it must be patchy (Stern et al
1995; \gierlinski\ et al., 1997).

\subsection*{Patchy static magnetic corona}

The solid angle subtended by a disk below a (static) patchy corona
would be $\sim 2\pi$. Clearly this can be tested by looking at the
amount of reflection, which is easily seen in the LS spectra (e.g. 
Done et al., 1992).  Fitting simultaneously for the solid angle,
ionization state and relativistic smearing gives that the reflector
subtends a solid angle of $<2\pi$, is
predominantly neutral, and significantly broadened but {\it not} by as
much as expected if the reflecting material extended down to the last
stable orbit (\zycki\ et al., 1997; 1998; 1999; Done \& \zycki\
1999). This means that fitting without including the effects of
relativistic smearing and/or ionization does not lead to large
distortions of the derived parameters. Fitting only for solid angle
and ionization (\gierlinski\ et al., 1997; Zdziarski, \lubinski\ \&
Smith 1999) or for solid angle and smearing (approximated by a
Gaussian: Gilfanov et al., 1999; Revnivtsev et al.,
2000), show overwhelmingly that the solid angle is
significantly less than $2\pi$, correlated with the
spectral index in the sense that harder continuua show smaller
reflected fractions (Zdziarski et al., 1999; 
Gilfanov et al 1999; Revnivtsev et al 2000), {\it and} a
smaller amount of smearing of the spectral features (Gilfanov et al
1999; Revnivtsev et al 2000).

These correlations extend to the time domain. Frequency resolved
spectra show that the fastest variability is characterised by a 
flatter spectrum and lower reflected fraction (Revnivtsev et al.,
1999). Also, the characteristic break frequency in the
LS variability power spectra {\it increases} as the spectrum 
steepens and the reflected amplitude and
smearing increase (Gilfanov et al 1999; Revnivtsev et al 2000). 

The static patchy corona model above a mainly neutral disk is ruled
out by the observations of $\Omega/2\pi<1$ for hard spectra (see
Beloborodov 2001, these proceedings). For magentic coronae
models to work then either the coronae are not static or the disk is
ionized.

\subsection*{Outflowing magnetic corona}

Outflow from magnetic reconnection is seen in the Sun, 
so it could also happen above the disk.  If
the outflow velocity is transrelativistic then the hard X--ray
radiation from the reconnection is beamed away from the disk. This reduces
the disk illumination, and thus the reflected fraction and the
reprocessed soft seed photons. Faster outflow velocities then give
fewer reprocessed soft photons and smaller reflected fractions (giving
the $\Gamma-\Omega/2\pi$ correlation), and if these faster velocities
are concentrated towards smaller radii then this would also suppress
the relativistic smearing signature from the inner disk (Beloborodov
1999). 

However, the mechanism for the spectral transition (LS--HS) is not
entirely clear. Given that it seems likely that the HS/VHS emission is
powered by reconnection, then it requires that the spectral transition is
associated with the magnetic loops mainly reconnecting 
inside the optically thick
material, rather than buoyantly rising above the disk. One potential
trigger for this could be that the 
disk becomes radiation pressure dominated -- although AGN disks
are in general radiation pressure dominated yet can show hard spectra
and low reflected fractions e.g. Zdziarski et
al., (1999); Chiang et al., (2000); Done et al.,
(2000). Similarly the link to the variability power spectra is also
not obvious: the particular radius picked out by the QPO/low frequency
break cannot be the radius at which the disk becomes radiation
pressure dominated as this would {\it increase} as the luminosity
increases from LS--VHS while the power spectral features imply a {\it
decreasing} radius.

\subsection*{Patchy corona above an ionised disk: I}

This seems immediately ruled out by the observation that the reflected
spectrum is mostly neutral in the LS. However, one way to suppress both
line and reflected continuum is from a radial dependance of ionization
state. If the inner region of the disk were so highly ionised that
even iron is completely stripped then it produces no atomic features.
Its reflected continuum is unobservable in the 2--20 keV range
and it appears instead to be part of the power law continuum.
The observed reflected spectrum could then come from material further
out in the disk which is mostly neutral, and not highly
relativistically smeared. Since the reflected fraction is the ratio of
the observed reflection relative to the inferred incidence flux
then $\Omega/2\pi$ is low both due to the inner disk solid angle being
missed and because this unidentified reflected signature boosts the
incident flux (Ross et al., 1999)

However, for a smooth (power law) distribution in ionization parameter
with radius then between the extreme ionization inner disk and more
neutral outer disk there must be an intermediate region of high
ionization, where iron is mainly in H-- and He--like ions. These
produce huge spectral features (see figure 3, right panel), which are 
unmistakable. The ionised disk models can be split into radial zones
of different ionization to test these ideas and in Cyg X--1 they can
be shown to be incompatible with the data (Done \& \zycki\ 1999).
Admittedly these models do not properly model the Compton upscattering
of the edge features (and slightly underestimate the Compton broadening of the
line: Ross et al., 1999) associated with the high ionization
material, but this is probably not a large distortion.

Similarly, the assumption of a constant ionization parameter slab can
(and should) break down in the vertical direction. Illumination of the
surface of the disk can result in an extreme ionization skin, which is
completely reflective. Some of the incident flux is reflected in the
skin, reducing the irradiating flux in the lower layers, so their
ionization state is smaller. Deep enough in the disk then the
material can be mostly neutral, giving the observed reflected
signature which has $\Omega/2\pi < 1$ as the illuminating flux for
that layer is not the full incident flux and the apparent incident
flux is boosted by the featureless reflected flux from the extremely
ionised upper skin (Ross \& Fabian 1993; Ross et al., 1999). 
Again though, there are problems with the
intermediate ionization layers. If the ionization parameter varies
smoothly as a (power law) function of height in the disk then between
the extreme ionization and mostly neutral material there should be the
highly ionised H-- and He-- like iron, with its unmistakable
reflected spectral features. However, Comptonization is now much more
of a problem as this spectrum has to be transmitted upwards through
the extremely ionised, hot skin. This additional Compton broadening
{\it may} be able to smear these features enough to disguise their
presence.  Such models have yet to be tested against the data but it
seems likely that if the Comptonization is strong enough to distort
the H-- and He-- reflected spectrum then it should also distort the
observed {\it neutral} reflection signature. This also has to be
transmitted through (even more of) the Comptonizing layers, and yet it
is easily detected.

Again in these models, as with the outflowing corona, there is no
obvious trigger for the LS--HS transition or particular radius to use
for the power spectral variability.

\subsection*{Patchy corona above an ionised disk: II}

A far more wideranging problem with the ionised reflection models
considered to date is that they {\it assume} a given density structure
as a function of height. At best, the codes take this structure
and then work out the temperature and ionization state
of the material as a function of height. But 
X--ray illumination {\it changes} the
density structure of the material. Material at the top of the disk is
heated by the illumination, and so can expand, lowering its
density. Deeper into the disk there is less X--ray heating, so the
material is cooler, and hence denser. 

The self--consistent density
structure is especially important as there is a thermal ionization
instability which affects X--ray illuminated material in pressure
balance (Field 1965; Krolik, McKee \& Tarter 1981; Kallman \& White
1989; Ko \& Kallman 1994; \rozanska\ \& Czerny 1996; Nayakshin, Kazanas
\& Kallman 2000, hereafter NKK). The disk is in hydrostic equilibrium
in the vertical direction, so the gas pressure, $P_{gas}$ must increase with
depth into the disk. In these circumstances, it is clearer to talk
about the ionization structure in terms of the pressure ionization
parameter $\Xi=P_{rad}/P_{gas}$ where $P_{rad}= L/(4\pi r^2 c)$ is the 
radiation pressure of the illuminating radiation (this relates to the
density photo--ionization parameter $\xi\sim \Xi kT/(4\pi$). 

The X--ray heating clearly depends on height, with the material at the
top of the layer being heated the most.  Because the material is in
pressure balance, the gas pressure is smallest at the top of the disk,
so $\Xi$ is at its maximum here. This corresponds to the uppermost
stable branch of the ionization equilibrium S--curve on an $\Xi-T$ plot,
with low density material which is highly or completely
ionized. Atomic cooling is negligible so the temperature is a fraction
of the local Compton temperature (see Krolik et al., 1981 and
Nayakshin 2000); Compton heating is balanced by Compton cooling and
bremsstrahlung. Going deeper down into the disk, the X--ray heating
decreases with optical depth to electron scattering, $\tau_T$, and the
gas pressure increases. Hence, both the temperature and ionization
parameter $\Xi$ decrease. However, there is then a turning point on
the S--curve, where the temperature and density
are such that the rapidly increasing bremsstrahlung cooling causes the
temperature to dramatically decrease. This pulls the gas pressure
down, but hydrostatic equilibrium requires that the gas pressure must
monotonically increase as a function of depth into the disk.  The only
way for the pressure to increase in a region with rapidly decreasing
temperature is to rapidly increase the density, but this pushes up the
cooling still further, and so decreases the temperature. Ionic and
atomic species can now exist, so line transitions give a yet further
increase in the cooling. Eventually this stabilises onto the bottom
part of the S curve, where the X--ray heating is balanced primarily by
line cooling from low temperature, high density material. This results
in an almost {\it discontinuous} transition from a highly ionized, hot
and relatively tenuous skin to mainly neutral, cool and relatively
dense material.

Reflection can then be suppressed as in the previous section, by the
decrease in illuminating flux on the neutral layer because of
scattering in the ionised skin, and because the scattered (reflected)
spectrum from the ionised skin is interpreted as incident continuum.
The difference is that the ionization parameter can now make an {\it
abrupt} rather than smooth transition from highly or extremely ionised
to mostly neutral material (NKK).

The resulting reflected spectrum is crucially determined by the optical depth
where this instability occurs {\it and} the Compton temperature of the
skin (which depends on the illuminating power law as well as the local
thermal disk spectrum).  For hard spectra, $\Gamma\sim 1.6$, then the
Compton temperature of the upper layer of the disk is high $\sim 10$
keV, its ionization is extreme so it produces no reflected spectral
features. If the instability occurs after an optical depth of
$\tau\sim 1$ then the apparent reflected fraction is $\sim 0.2$ from
neutral material, as observed.  For steeper spectra the Compton
temperature of the upper layers drops. Thus the instability point
occurs rather higher in the disk, so the optical depth of the skin
decreases, and the amount of reflected flux increases (giving the
$\Gamma-\Omega/2\pi$ correlation! NKK). 

The right panel of Figure 1 shows results from fitting simulated 2--20
keV GINGA resolution spectra of these full vertical structure models
with the fixed density reflection models. The apparent
$\Gamma-\Omega/2\pi$ is clearly seen for the LS $\Gamma < 2$ spectra
(Done \& Nayakshin 2000).

These models concentrate on the vertical ionisation structure, but
there should also be a radial dependance. Larger radii should be less
ionised and the total disk spectrum is found by integrating over
all these different ionisations. However, the meaning of less ionized
is very different from that in the constant density models -- the skin
remains completely ionized with a negligible amount of H and He--like
iron, but its Thomson depth decreases with radius (see Nayakshin
2000). Thus the inner disk can have a larger skin depth, so
suppressing the most extreme relativistically smeared line
components (Nayakshin 2000). 
Perhaps this also gives a radius where the skin depth changes
from $\tau_T>1$ to $\tau_T <<1$ which could pick out a characteristic
frequency for the power spectral variability, although the mechanism
for the LS--HS transition is again not clear (see previous section).

\subsection*{Truncated disk with inner X--ray hot flow}

All the observations can all be qualitatively
explained in the truncated disk--advective inner flow model, with the
disk progressively penetrating further into the advective flow. The
geometry automatically gives a rather `photon starved' hard continuum.
As the overlap between the two phases increases, the 
soft seed photon flux intercepted by the hot inner flow gets larger, 
and so leads
to a steeper continuum. But the increasing overlap 
also gives a larger solid angle subtended by the
disk (resulting in a $\Gamma-\Omega/2\pi$ correlation: Poutanen, Krolik \&
Ryde 1997). The decreasing inner radius of the disk also gives the
increasingly relativistically smeared spectral features, {\it and}
higher frequency power spectral features. The advective flow collapses
at luminosities at a few per cent of Eddington, giving a physical
mechanism for the state transition which is observed at these fluxes.

The only problem is a quantitative one. The inner disk radii deduced
from the relativistic smearing are $\sim 10-20$ Schwarzchild radii.
Standard advective flow calculations assume that the transition radius
between the SS disk and hot inner flow is large $\sim 10^4$
Schwarzchild radii. This is an assumption, and can be changed.  The
transition radius can be physically determined by calculating the
evaporation rate from the ionised disk skin (\rozanska\ \& Czerny 2000), but
irrespective of the mechanism, moving the disk inwards (and especially
increasing the overlap between the two flows) increases the soft seed
photon flux intercepted by the advective flow, leading to stronger
Compton cooling. Bringing the disk in to $\sim 10-20$ Schwarzchild
radii lowers the maximum luminosity at which the advective flow can
exist by a factor of 2--3 (Esin 1997), which then makes the LS--HS/VHS
state transition at slightly lower luminosities than observed.

However, the ideas developed above about the ionization structure of
the disk should also apply here. Any hydrostatic disk illuminated by
hard X--rays is subject to the thermal instability, and so should have
an ionised skin. The observed neutral reflected spectrum is
transmitted through this skin, and is subject to Compton
broadening. If some fraction of the width of the line is due to
Compton scattering rather than relativistic effects then this would
result in an underestimate of the disk radius, and so extend the
maximum luminosity of the advective flow back into the observed range.

\section*{COMPARISON WITH OBSERVATIONS OF GBHC: HIGH AND VERY HIGH STATE}

Much less work has been done on X--ray reflection in the soft states.
The only spectra which have been analysed with both ionisation and
relativistic smearing included in the reflected spectrum are Nova Muscae
(\zycki\ et al., 1998), RXTE J1550-564 (Wilson \& Done 2000) 
and Cyg X--1 (Gilfanov et al., 1999; \gierlinski\ et al
1999). Fitting with the fixed
density/ionization structure models gives a solid angle of $\sim
0.2-1$ of strongly ionised material, and that the spectral
features are strongly smeared. Fits which do not include ionisation of
the reflector will {\it overestimate} the solid angle subtended by
the reflecting material (compare the left and right panels of 
Figure 3), so the results of 
Gilfanov et al., (1999) on the $\Gamma-\Omega/2\pi-$ smearing
correlations will probably change significantly for the steep spectrum
objects in their sample when ionisation is included. 

The true solid angle is again hard to infer due to ionization effects.
Even with the fixed density reflection models it depends on how the
ionization varies as a function of radius (\zycki\ et al.,
1998). Obviously the ionization can also vary as a function of height,
and the ionization instability again should operate. For steep
spectral illumination this does not generally lead to suppression of
the observed reflected flux as the Compton temperature of the skin is
not high enough to completely strip iron. Instead it is dominated by
H-- and He--like iron and so its reflected spectrum can be seen
directly (Figure 1, right panel; Done \& Nayakshin 2000). But yet
another problem for the reflection models is that the disk should now
be substantially collisionally ionised, as well as photo--ionised.

Even more uncertainty comes from the geometry.  If the
thermal/non--thermal electrons are co--spatial then both should
illuminate the disk. But if only the non--thermal electrons are
associated with magnetic reconnection, while the thermal electrons are
from a heated disk skin then most of the incident flux for reflection
will be only from the non--thermal power law, while the observed
incident continuum below 20 keV can be dominated by the thermal
Compton component. This would dilute the observed solid angle (Wilson
\& Done 2000; \gierlinski\ et al., 1999), but the Compton continuum
produced from a non--thermal electron distribution can be highly
anisotropic, which could boost the observed solid angle (Ghisellini et
al., 1991)!

\section*{CONCLUSIONS}

GBHC show different spectral states which roughly correspond to
luminosity, with the LS--HS/VHS transition occurring at 2--4 per cent
of Eddington.  In the LS ($1.5 < \Gamma < 1.9$), 
fitting the spectra with (fixed density)
models of X--ray reflection gives that 
the solid angle subtended by the disk is small $\Omega/2\pi <1$, that
the material is mainly neutral, and that the spectral features are
broadened, but not by as much as expected from relativistic effects if
the disk extended down to the last stable orbit around the black
hole. If the smearing is due to relativistic effects alone then it
implies that the reflecting material only extends down to $10-20$
Schwarzchild radii. All these correlate, with steeper spectra having
more reflection and more broadening (and higher frequency features in
the power spectra).

These observations rule out the continuous corona model (from the
overall hard spectral shape) and a static patchy corona above a
neutral disk (from $\Omega/2\pi < 1$) and a static patchy corona above
an ionised disk (as the reflector is mainly neutral)
and advective flows (from the combination of small inner radius and
high luminosity). Only the outflowing corona (Beloborodov 1999) can
fit these `observed' parameters. 
Radial ionization structure is unlikely to change these conclusions, 
although vertical ionization structure can give some
loopholes due to Compton scattering of the observed reflected
signature (Ross et al., 1999). 
This {\it may} distort the reflected signature from a
static patchy corona by enough to not see the high ionization zone,
and Compton broadening {\it may} 
allow the advective flow inner radius to be significantly 
larger. 

However, the constant density models can be seriously misleading due
to the ionization instability (NKK). This can cause a discontinuous
transition between an extremely ionised skin overlying mostly neutral
material, and can reproduce the observations (and correlations) with a
static patchy magnetic corona. 
But this instability should be present
whatever the mechanism for illumination of the disk, so it can also
co--exist with either an advective flow or an outflowing corona. This
of course completely removes any diagnostic power from the fixed
density reflection models (Done \& Nayakshin 2000). 

For the HS and VHS the reflected spectrum is
plainly highly ionised and broadened. However, the solid angle is even less
constrained than in the LS due to the complex continuum, which could
represent a single thermal/non--thermal electron population or could
represent two entirely spatially distinct regions, with different
solid angles for illumination of the disk material. 

Reflection {\it may} be able to live up to its promise of illuminating the
geometry (and so distinguishing between various ideas as to the 
underlying emission mechanisms)
but {\it only} with much more detailed modelling of the X--ray
irradiated disk. 

\section*{ACKNOWLEDGEMENTS}

Much of this review includes ideas developed in collaborations with S.
Nayakshin, C. Wilson and P. \zycki.

\end{document}